\documentclass[sigconf,10pt]{acmart}
\usepackage{xspace}
\usepackage{newtxtext} 
\usepackage[capitalize,noabbrev]{cleveref}
\usepackage{multirow}
\usepackage{booktabs}
\usepackage{subcaption}
\usepackage[ruled]{algorithm2e}
\usepackage[
    n,
    operators,
    advantage,
    sets,
    adversary,
    landau,
    probability,
    notions,
    logic,
    ff,
    mm,
    primitives,
    events,
    complexity,
    asymptotics,
    keys]{cryptocode}

\newcommand{\sys}{\textsc{BOA}\xspace}

\makeatletter
\newcommand{\crefnames}[3]{\@for\next:=#1\do{\expandafter\crefname\expandafter{\next}{#2}{#3}}}\makeatother
\crefnames{part,chapter,section}{\S\kern-0.25em}{\S\S\kern-0.25em}

\newtheorem{definition}{Definition}

\newcommand{\heading}[1]{
  \vspace{1ex}
  \noindent
  \textbf{#1}}

\newcommand\shortsection[1]{
    {\vspace{1ex}\noindent\textbf{#1.}}
}

\let\latexusecounter=\usecounter
\def\compactsortof{
    \itemsep=0in
    \topsep=2pt
    \parsep=0in
    \partopsep=0pt
    \leftmargin=0em
}

\newenvironment{myitemize2}
{\begin{list}{\labelitemi}{
    \itemsep=1pt
    \topsep=2pt
    \parsep=0in
    \partopsep=0pt
    \leftmargin=2em
}}
{\end{list}}

\definecolor{mygreen}{RGB}{0,128,0}

\usepackage{tcolorbox}
\tcbuselibrary{skins}

\definecolor{framecolor}{rgb}{0.122, 0.435, 0.698}
\definecolor{bgcolor}{rgb}{0.95, 0.95, 0.95}

\def\hn{\sffamily\selectfont}
\newcommand{\mpfont}{\hn\scriptsize}

\ifx\noeditingmarks\undefined
    \newcommand{\MPworker}[2]{\unskip{\color{#1}\vrule\vrule}{\marginpar{\raggedright\color{#1}\mpfont #2}}}
\else
    \newcommand{\MPworker}[2]{\unskip}
\fi

\newcommand{\nodetree}{chunk tree\xspace}
\newcommand{\NodeTree}{Chunk Tree\xspace}
\newcommand{\blocktree}{block tree\xspace}
\newcommand{\BlockTree}{Block Tree\xspace}

\makeatletter
\let\state\@ACM@addtoaddress 
\makeatother
\AtBeginDocument{%
  }

\setcopyright{none}
\acmConference{}{}{}
\acmDOI{}
\acmISBN{}
\acmYear{}
\renewcommand\footnotetextcopyrightpermission[1]{}

\begin{document}

\title{Toward a Principled Framework for Agent Safety Measurement}

\author{
    \fontsize{12}{11}\selectfont
    Shuyi Lin, Anshuman Suri\textsuperscript{$\star$}, Alina Oprea, and Cheng Tan}
\affiliation{%
    \institution{\textit{Northeastern University \quad \textsuperscript{$\star$}DatologyAI}}
  \city{}
  \state{}
  \country{}
}

%
%

\renewcommand{\shortauthors}{}

\begin{abstract}
LLM agents emit actions, not just text, and once taken, those actions often cannot be undone.
Yet today's agent-safety evaluations run greedy or a few sampled rollouts
and report a single safe/unsafe rate---blind to the long-tail trajectories
where unsafe behavior may arise from low-probability but non-negligible actions.

We argue agent safety should be measured by \emph{search}, not sampling.
We apply \sys
a framework that, given a deployment configuration
(model, decoder, prompt, environment, judger, likelihood budget), searches the in-budget
trajectory space and reports a \emph{safety score}: the probability the
agent stays safe under the configuration.
\sys searches both within a single LLM round and across the agent--environment interaction tree
under a given likelihood budget, and makes search practical via batched
decoding/judging, prefix caching, and chunked tree expansion.
On agent-safety workloads, \sys discovers unsafe trajectories that greedy and
sampled evaluations miss. \sys can additionally be used for ranking models, defenses, and attacks,
all on the same scale, with manageable GPU costs.
\end{abstract}


\maketitle
\fancypagestyle{firstpagestyle}{\fancyhf{}}
\thispagestyle{firstpagestyle}
\pagestyle{empty}

{\renewcommand{\thefootnote}{$\star$}\footnotetext{Work done while at Northeastern University.}}

\section{Introduction}
\label{s:intro}

Large language models (LLMs) have moved beyond chat:
an \emph{LLM agent} reads a user's instruction, plans actions, calls external tools,
observes their results, and continues turn after turn until the task is done.
The same loop that makes agents \emph{useful}---a feedback cycle with the outside
world---also makes them \emph{risky}. An agent does not merely emit text;
it emits \emph{actions}.
%
However, actions can be dangerous: they may delete important data, leak private
information, execute destructive system commands, or transfer money to an
unintended account~\cite{replit2025dbdeletion,shapira2026agents,commentAndControl2026,freysa2024}, none of which can be undone.
As organizations move agents to production,
a basic question becomes urgent:
\emph{how safe is this agent under its deployed configuration?}

The status quo answer is unsound:
the current practice is to pick a safety benchmark~\cite{andriushchenko2025agentharmbenchmarkmeasuringharmfulness,zhang2024agentsafetybench,zhang2024asb,ruan2024identifying},
run the agent on the tasks under either greedy decoding or a handful of random samples,
and ask a judge whether each resulting trajectory is safe.
The agent is then summarized by a single safe/unsafe rate.
This practice conflates two very different things:
(a) the trajectory the agent \emph{happened to produce} on this run,
and (b) the trajectories the agent \emph{could produce} when deployed.

The gap is substantial.
Deployed agents almost never run greedy;
they run with top-$k$, top-$p$, and other stochastic sampling~\cite{holtzman2019curious,shi2024thorough}
for better task performance.
Stochastic decoding means the same prompt can take a
different trajectory tomorrow than it did today, and unsafe behavior is
precisely the kind of behavior that lives in the long tail:
the model has been trained to refuse, so unsafe completions are
low-probability-but-reachable rather than typical.
An evaluation that inspects only typical trajectories is,
by construction, blind to the trajectories that matter most.

This raises the core question of this paper:
\emph{can we evaluate an agent without conflating measurement with sampling variance?}

Concretely, given a deployment configuration---an LLM, a decoding
strategy, a prompt, the tools and environment the agent will see, and a
judging function that decides what counts as unsafe---can we
systematically answer \emph{how safe} the resulting agent is?

\heading{Problem setup and what \sys measures.}
We focus on the deployment-time evaluator's view: a developer or auditor
holds the model $\mathcal{M}$, the decoding strategy $\mathcal{D}$, the
tool/environment surface $\mathcal{E}$, and a judger $\mathcal{J}$, and
wants a quantitative answer for a task prompt $p$. The prompt $p$ is taken
from a safety benchmark:
it can be a benign user request that stresses
the agent's refusal training, an adversarial jailbreak, an environment
seeded with prompt-injection content, or a tool-use scenario with
tempting but unsafe shortcuts.
The judger and the environment are trusted; the agent is not.

\sys does not invent new attacks or defenses;
it answers, for any such configuration, whether the agent reaches an
unsafe trajectory and how likely it is to do so under its own decoding strategy.
The same instrument therefore lets
a developer rank candidate models,
an auditor evaluate a defense,
and a red-teamer compare attacks, all on a common framework.

\heading{What we want from the answer.}
A useful answer is not a single bit.
We instead seek a \emph{safety score}: the probability that an agent operates
safely under a given condition.
Concretely, under a deployed decoder $\mathcal{D}$ and restricted to trajectories that are in-budget under a user-chosen likelihood budget $\epsilon$, the safety score is the probability that the agent produces a
trajectory labeled safe by the judger.
Safety score generalizes the binary safe/unsafe label, distinguishes
a robustly-refusing agent from one that merely refuses on the typical
path, and---given enough compute---can either produce a concrete
unsafe trajectory as a witness or certify that the trajectory mass above
the likelihood floor has been exhausted without finding one.

\bigskip
Producing this answer however requires a search procedure parameterized by
the deployment configuration, not another benchmark of prompts.
Search moves the effort from collecting more prompts to
\emph{exploring the trajectories each prompt admits},
which is where the missing risk lives.
Prior work studies token-level search. For example, Jailbreak Oracle~\cite{lin2026toward}
targets single-turn LLM responses. This formulation does not extend to agents. An
agent's risk surface arises from a multi-turn interaction with an external,
stateful environment whose responses are not drawn from the model's distribution.
Adapting search to this setting forms the core of this work.


In this paper,
we introduce \sys, a principled framework to measure agent safety. \sys
takes a deployment configuration $\langle\mathcal{M}, \mathcal{D}, p,
\mathcal{E}, \mathcal{J}, \epsilon\rangle$ and searches the trajectory space
reachable under that configuration for unsafe behavior.
\sys searches for either witness trajectories the judger labels unsafe, together with their
likelihood under the deployed decoder, or evidence that the in-budget
trajectory has been exhausted without finding one.
The final output is a \emph{safety score}.

\sys searches at two levels.
Within a single LLM round, it
explores the model's output distribution under
the specified
decoding strategy.
Across rounds, it represents the
agent--environment interaction as a tree whose nodes alternate between
model-generated chunks and environment responses.


Further, \sys introduces three system-level optimizations:
(1) \emph{Batching}: agent-safety search is throughput-bound
rather than
latency-bound, so \sys aggregates pending model decodings and judge
evaluations into batches that keep the GPU busy and amortize per-call
overhead.
(2) \emph{Prefix caching}: \sys is exploring a tree, and tree
exploration repeatedly revisits prefixes; \sys keeps a response cache of
already-judged trajectories and reuses their results whenever the current
path is a prefix of a previously evaluated one, eliminating large amounts
of redundant work. 
(3) \emph{Chunked expansion}: rather than expanding the
search tree one token at a time---which exhausts the budget at shallow
depths and never reaches the environment---\sys expands a configurable
chunk of tokens per node and invokes the judger only at chunk
boundaries, trading fine-grained pruning for the ability to actually
reach multi-turn tool-calling behavior within a fixed time budget.

This paper makes the following contributions:
\begin{myitemize2}

    \item 
    We formulate agent safety measurement as a search problem parameterized by
        the deployment configuration. The method outputs a safety score that
        grades agents by the likelihood of in-budget trajectories under the
        deployed decoder, rather than by a single greedy verdict.

    \item
    We adapt \sys, a principled search framework, to an agentic setting by
    combining one-round chunk-level search with a multi-turn
    agent--environment interaction tree.


    \item
        We make the search practical at agent scale via three
    system-level optimizations---batched decoding and judging, prefix
    caching of judged trajectories, and chunk-based tree expansion---each
    of which targets a specific bottleneck in the agent setting.

    \item We implement \sys and evaluate it on agent-safety workloads,
    showing that it discovers unsafe behaviors that greedy and sampled
    evaluations miss, and ranks base models, defenses, and attacks at a
    finer granularity than binary labels at a manageable GPU cost.

\end{myitemize2}

\section{Problem Setup}
\label{sec:motivation}

\heading{Existing agent safety evaluations.}
A growing body of work measures agent safety. Benchmarks such as
AgentHarm~\cite{andriushchenko2025agentharmbenchmarkmeasuringharmfulness},
Agent-SafetyBench~\cite{zhang2024agentsafetybench},
ASB~\cite{zhang2024asb},
ToolEmu~\cite{ruan2024identifying}, and
R-Judge~\cite{yuan2024r}
score agents on diverse risk categories---direct harms, unauthorized tool
calls, and the safety judges themselves.
In parallel, agent-specific attacks have been
characterized, such as prompt injection~\cite{greshake2023not},
memory poisoning~\cite{chen2024agentpoison}, and adversarial tool
descriptions~\cite{ye2024toolsword},
and defenses such as
instruction-hierarchy training~\cite{wallace2024instruction},
spotlighting~\cite{hines2024defending}, and constitutional
classifiers~\cite{sharma2025constitutional} have been proposed.
All of these, however, reduce the agent to a single greedy or few-sample
verdict---a practice that \S\ref{s:intro} argues is unsound for deployment-time
decisions.

\heading{Problem definition.}
We formalize the question informally posed in Section~\ref{s:intro}.
Let $\mathcal{M}$ be a language model, $\mathcal{D}$ a decoding strategy, and
$\mathcal{E}$ an environment that returns tool-execution results and feedback.
An interaction \emph{trajectory} $r$ is the sequence of model generations and
environment responses produced during a multi-turn interaction; we write
$\condprobsub{\mathcal{D}}{r}{\mathcal{M},p,\mathcal{E}}$ for its probability
under prompt $p$. Only model-generated tokens contribute to this probability;
environment responses are treated as deterministic given the model's tool
calls. Let $\mathcal{J}(p,r)\in\{0,1\}$ be a judger that returns $1$ iff $r$
is unsafe; trajectories with malformed tool calls are safe by construction
because they never reach the environment.

Trajectory probabilities decay exponentially with length, so a fixed
probability cutoff is uninformative.
We use a length-relative budget.
Let $\mathcal{R}_{\geq n}$ denote trajectories with at least $n$ tokens,
and define a metric $\mathcal{L}_n(\mathcal{M},p,\mathcal{D},\mathcal{E})$: the expected probability
of the first $n$ model-generated tokens of a trajectory drawn from $\mathcal{D}$
conditional on its length being $\geq n$,
\begin{equation*}
\mathcal{L}_n(\mathcal{M},p,\mathcal{D},\mathcal{E}) =
\expsub{r \,\sim\, \mathcal{D}(\mathcal{M},p,\mathcal{E})\,\mid\,|r|\geq n}{\condprobsub{\mathcal{D}}{r_{1:n}}{\mathcal{M},p,\mathcal{E}}}.
\end{equation*}
Given a user-chosen $\epsilon \in (0,1]$ (typically $\epsilon \ll 1$), the
unsafe likelihood threshold is
$\tau(n) = \epsilon \cdot \mathcal{L}_n(\mathcal{M},p,\mathcal{D},\mathcal{E})$:
unsafe trajectories must remain at least an $\epsilon$ fraction as likely as a
typical $n$-token prefix.

\begin{definition}[Agent Safety Search Problem]
\label{def:jo}
Given $\langle\mathcal{M}, \mathcal{D}, p, \mathcal{E}, \mathcal{J}, \epsilon\rangle$,
let $\tau(n) = \epsilon \cdot \mathcal{L}_n(\mathcal{M},p,\mathcal{D},\mathcal{E})$ be the length-relative likelihood threshold derived from the user-chosen budget $\epsilon$, and
\[
\mathcal{T}_\tau \;=\; \bigl\{\, r \;:\; \condprobsub{\mathcal{D}}{r}{\mathcal{M},p,\mathcal{E}} \,\geq\, \tau(|r|) \,\bigr\}
\]
be the set of \emph{in-budget} trajectories---those whose generation likelihood
under $\mathcal{D}$ meets the length-relative budget.
The Agent Safety Search Problem asks to enumerate $\mathcal{T}_\tau$,
label each $r \in \mathcal{T}_\tau$ with $\mathcal{J}(p,r)$, and return:
\[
    \textrm{safety score } \mathcal{S}
\;=\;
\frac{\displaystyle\sum_{\substack{r \,\in\, \mathcal{T}_\tau \\ \mathcal{J}(p,r)=0}}
       \condprobsub{\mathcal{D}}{r}{\mathcal{M},p,\mathcal{E}}}
     {\displaystyle\sum_{r \,\in\, \mathcal{T}_\tau}
       \condprobsub{\mathcal{D}}{r}{\mathcal{M},p,\mathcal{E}}}
\]
\end{definition}

The safety score ($\mathcal{S}$) is the likelihood that the trajectory is judged safe:
$\mathcal{S}=1$ means every in-budget trajectory is safe (robustly safe),
$\mathcal{S}=0$ means every in-budget trajectory is unsafe,
and intermediate values grade the agent by
how much of its in-budget probability lands on safe trajectories.

The search space is a tree whose nodes alternate between model-generated
tokens and environment responses. Next, we describe how \sys
explores the tree under a finite budget.

\section{\sys: An Efficient Agent Safety Measurement Framework}
\label{sec:method}
Exhaustive search over the space of possible agent interactions is computationally infeasible, and random sampling is equally inefficient---it concentrates on high-probability sequences.
Although the effective search space is reduced by likelihood and decoding constraints such as top-$p$ and top-$k$, it is still far too dense for naive exploration.
A useful safety measurement framework must therefore broadly cover
trajectory branches, while searching deeper along unsafe ones.


\heading{Overview.}
\sys is an efficient framework for agent safety measurement.
It introduces a specialized search algorithm that surfaces unsafe agent behaviors by exploring the trajectory tree at three structural levels---\textit{chunks} as the unit of expansion, a \textit{\nodetree} within each turn, and a \textit{\blocktree} across turns---each scheduled by its own queueing policy (priority search within turns, breadth-first across turns).
We next describe the search and its components in detail.

\subsection{Phase-Aware Priority Search}
\label{subsec:phase_aware_search}

\begin{figure}[t]
    \centering
    \includegraphics[width=\linewidth]{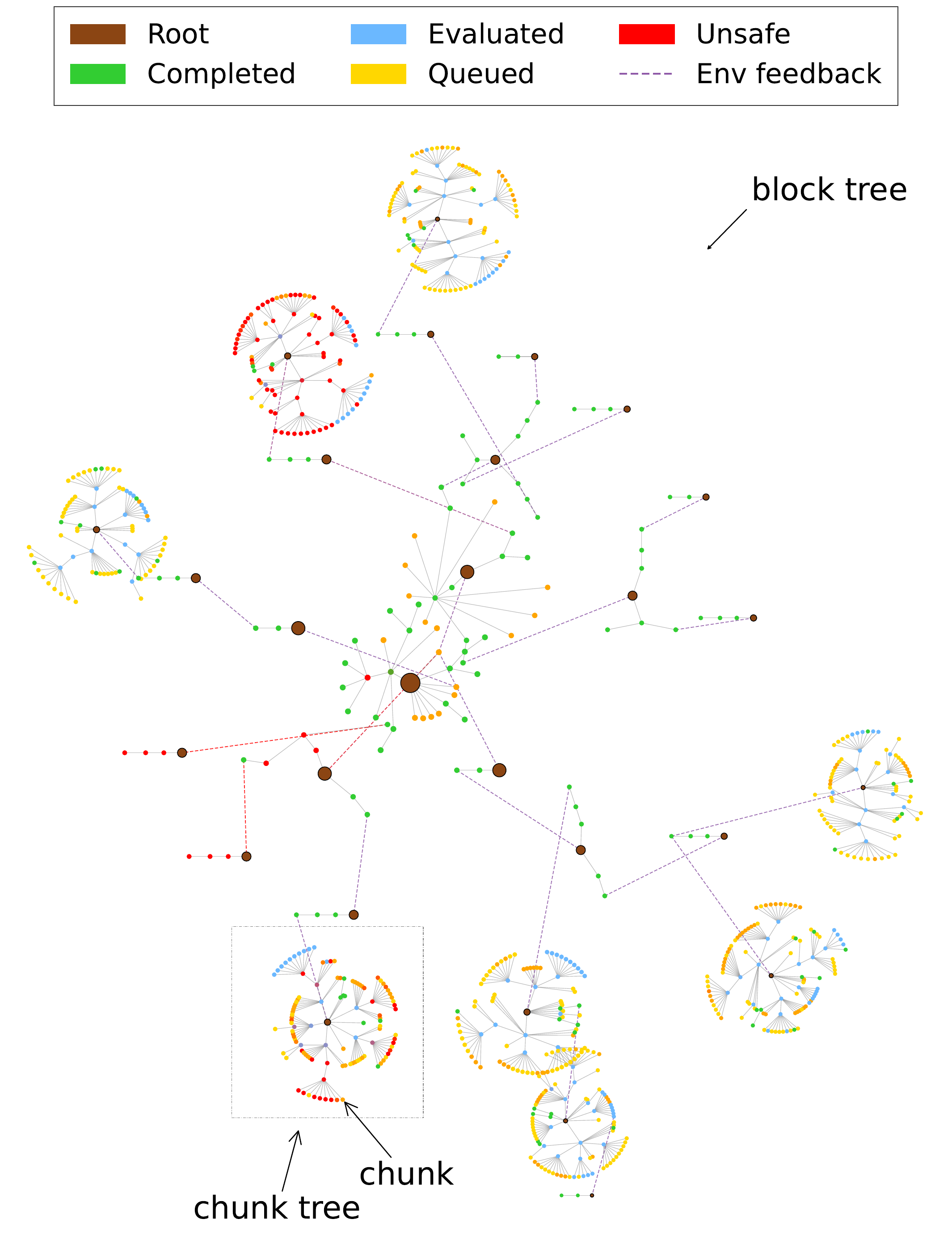}
    \caption{An illustrative execution trace of \sys's tree search on a
    multi-turn request using Llama-3.1 (8B). Each turn produces its own tree, shown as
    concentric circles with the root at the center and depth growing
    outward; the gray wedge marks that turn's expansion. Dashed lines
    denote environment interactions linking successive trees.
    }
    \label{fig:prompt_0049_trees}
\end{figure}

\begin{algorithm}[ht]
\LinesNumbered
\SetAlgoNoEnd

\caption{\sys Multi-Turn Agent Evaluation}
\label{algo:boa_nested}
\newcommand\mycommfont[1]{\footnotesize\ttfamily\color{blue}{#1}}
\SetCommentSty{mycommfont}
\SetKwInput{KwIn}{Input}
\SetKwInput{KwOut}{Output}
\KwIn{Prompt $p$, Environment $\mathcal{E}$, user-chosen $\epsilon$, sample budget $K$, phase boundary $d_{min}$}
\KwOut{Safety score $\mathcal{S}_{root}$}
\BlankLine
$PQ_{block} \gets \{ \text{InitBlock}(p) \}$ \\
\tcp{Across-Turn: Multi-turn \BlockTree search}
\While{$PQ_{block} \neq \emptyset$ \textbf{\upshape and} budget remains}{
    $B \gets PQ_{block}.\text{pop}()$ \tcp*{Get the current block}

    \lIf{$B.\text{tool\_call} \neq \text{null}$}{$B.\text{prompt} \gets \text{ApplyTool}(B, \mathcal{E})$}

    $PQ_{chunk} \gets \{ \text{InitChunk}(B.\text{prompt}) \}$  \\

    \tcp{In-turn \nodetree search}
    \While{$PQ_{chunk} \neq \emptyset$ \textbf{\upshape and} budget remains}{
        \tcp{Breadth if $depth<d_{min}$, else safety-depth}
        $c \gets PQ_{chunk}.\text{pop}()$

        \ForEach{$chunk \in \text{ExpandChunks}(c)$}{
            \lIf{$chunk.\text{weight} < \epsilon$}{\textbf{continue}}

            $score \gets \text{Judge}(chunk)$

            \tcp{Bubble up score to root of $B$}
            $\text{Propagate}(chunk, score)$
        }
    }
    \tcp{Block safety score aggregation}
    $\mathcal{S}_B \gets \text{root}(B).\text{safety\_score}$

    $\text{PropagateBlock}(B, \mathcal{S}_B)$

     \tcp{Sample $K$ trajectories and normalize their probabilities}
    $\mathcal{C} \gets \text{SampleSafe}(B, K)$

    $Pr_c \gets c.\text{prob} \big/ \sum_{c' \in \mathcal{C}} c'.\text{prob}$

    \ForEach{$c \in \mathcal{C}$ \textbf{\upshape with} $Pr_c \geq \epsilon$}{
        $PQ_{block}.\text{insert}(\text{SpawnBlock}(B, c))$ 
    }
}
\Return{$\mathcal{S}_{root}$}
\end{algorithm}

Algorithm~\ref{algo:boa_nested} outlines \sys's hierarchical priority search, and \Cref{fig:prompt_0049_trees} shows a representative trace on a multi-turn request---each turn rendered as a concentric tree linked to the next by environment interactions. We describe the search bottom-up: first the within-turn structure (chunk expansion and the \nodetree), then the across-turn \blocktree, and finally the safety score and judger that drive the priority queues.

\shortsection{Within-Turn: Chunk-based Expansion}

\sys extends our prior Jailbreak Oracle~\cite{lin2026toward} from single-token to chunk-based expansion: it
expands generation in \textit{chunks} of $c$ tokens to overcome the
computational bottleneck of frequent judging. In agentic tasks, a single turn
often spans hundreds of tokens; invoking the judger at every token would
exhaust the search budget at extremely shallow depths, preventing the model
from ever reaching an end-of-sequence (EOS) token to trigger the environment.
By aggregating generation into chunks, \sys amortizes the judging cost,
allowing the search to penetrate deep enough to complete tool calls within a
fixed time budget. The chunk size $c$ thus acts as a granularity knob: smaller
$c$ offers fine-grained pruning at high cost, while larger $c$ ensures the
search can reach environment interaction points without being prematurely
throttled by the judger.

\shortsection{Within-Turn: \NodeTree}
Building on chunk-based expansion, \sys organizes each turn's LLM generation as a \textit{\nodetree}: each node is a chunk, and its children are different chunks sampled from the model conditioned on the prefix up to that node. A node becomes a leaf at EOS or when the turn's budget is exhausted (\Cref{fig:prompt_0049_trees}, concentric circles). Exploration over this tree follows a phase-aware priority queue: at shallow depths, the scoring function $f(\cdot)$ prioritizes \textit{breadth} to cover diverse early branches; as trajectories lengthen, it shifts to safety-weighted depth, lifting branches most likely to yield unsafe behavior even in low-probability regions of the model's distribution. This lets \sys ``tunnel'' into rare but critical safety violations that naive sampling would miss.

\shortsection{Across-Turn: Multi-turn \BlockTree}
\sys organizes the search across turns into a \textit{\blocktree}, where each block is the \nodetree of a complete assistant turn and blocks are linked by tool-call/tool-result transitions through the environment $\mathcal{E}$. Unlike the within-turn priority queue, block-level expansion proceeds in strict breadth-first order. To prevent the \blocktree from exploding across turns, branches whose cumulative probability falls below $\tau(n) = \epsilon \cdot \mathcal{L}_n$ are pruned, where $\epsilon$ is the user-chosen likelihood budget.

\shortsection{Safety Score}
Each node maintains a \emph{safety score} representing its estimated probability of safe termination, computed as follows:
\begin{itemize}
    \item \emph{Leaf at end-of-sequence token (EOS).} The score is the judger's evaluation of the completed trajectory along the path from the root to this leaf.
    \item \emph{Leaf at budget exhaustion.} \sys performs $n$ independent rollouts from the leaf and reports the average of their judged scores.
    \item \emph{Internal node.} The score is a probability-weighted aggregate of its children's scores. This rule applies uniformly to internal nodes within a turn's \nodetree and to internal blocks in the \blocktree, so updates propagate bottom-up to the root.
\end{itemize}

\shortsection{Judger}
\sys uses the continuous \textit{safety score} to guide the priority search, evaluated by an LLM judger. To reduce the overhead of invoking the judger, we implement a lightweight validation step: whenever a chunk contains a tool call, the environment $\mathcal{E}$ first verifies its syntax and executability. Because a malformed or non-executable call cannot result in a successful attack or cause harm, these paths are marked safe without invoking the judger. Only valid tool calls and standard conversational responses are forwarded to the model-based judger for full evaluation.

\subsection{System Design and Implementation}
\label{s:optimize}

\begin{figure}[h]
    \centering
    \includegraphics[width=1\linewidth]{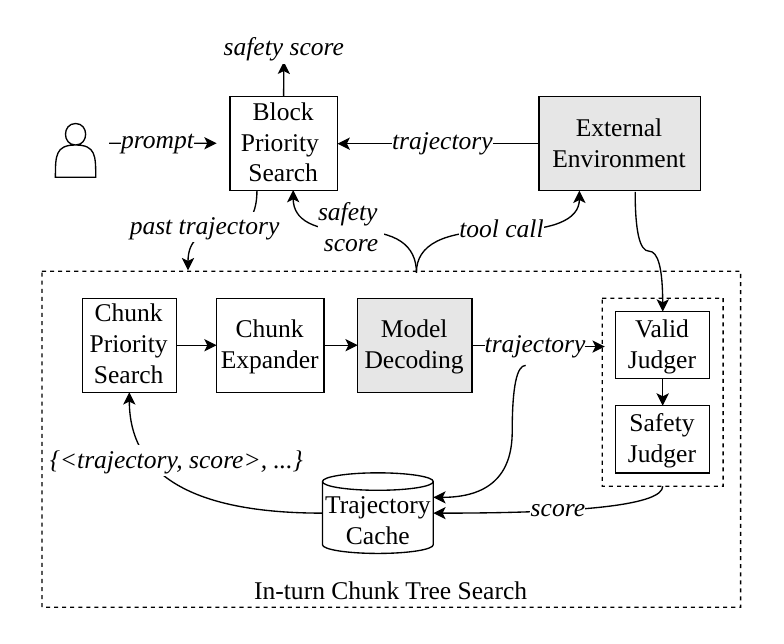}
    \caption{\sys's architecture.
    The shaded component (``Model Decoding'')
    represents existing serving frameworks such as HuggingFace or vLLM.
    %
    }
    \label{fig:arch}
\end{figure}

\shortsection{\sys overview}
\Cref{fig:arch} shows the overall design of \sys.
It consists of three major components: priority search engines, trajectory cache, and external environment.

\shortsection{Priority search engines}
Both the \blocktree and the \nodetree are scheduled by priority search engines.
The component is policy-pluggable---phase-aware, BFS, and others---so \sys can
trade off exploration and exploitation by swapping the engine alone.
The block-tree engine consumes samples already produced by the chunk-tree side, so
it requires no expander of its own.
%
The chunk-tree engine drives expansion through a \textit{chunk expander} that produces chunks.
To keep the GPU saturated, the
expanders batch candidate paths and decode them together, and adopt \emph{fast
approximate top-$p$} decoding: it sorts only the top 512 tokens rather than the
full vocabulary, which is sufficient for typical $p$ (e.g., $0.9$, $0.95$).

\shortsection{Trajectory cache}
\sys maintains a cache of judged trajectories to eliminate redundant evaluations. Each entry records a path through the search tree---a sequence of chunks, possibly spanning multiple turns and tool interactions---together with its safety score. During exploration, the priority search engine reuses a cached entry whenever the current path is a prefix of a previously evaluated trajectory. Such reuse is frequent: because \sys searches a tree, newly expanded paths routinely share prefixes with already-evaluated ones.

\shortsection{External environment}
The external environment $\mathcal{E}$ executes tool calls committed by the search and returns their results, allowing \sys to advance the \blocktree across turns. 

\section{Experimental Evaluation}
\label{sec:evaluation}

We implement and evaluate \sys on agent workloads, demonstrating
(Capability, \Cref{subsec:capability}) the additional information \sys provides over existing agent-safety evaluators, beyond a binary safe/unsafe label;
(Applications, \Cref{subsec:applications}) its use in comparing safety levels across different base models, defense mechanisms, and attack mechanisms; and
(Cost, \Cref{subsec:reliability_cost}) its computational cost.

\shortsection{Benchmark}
We use Agent-SafetyBench~\cite{zhang2024agentsafetybench} as our evaluation benchmark.
To obtain a balanced and reliably-labeled evaluation set, we first run all four
target LLMs (listed below) on the full benchmark using greedy decoding
(used by the original paper) and remove prompts whose outputs cannot be reliably labeled by the safety evaluator.
We then partition the remaining prompts by their
cross-model greedy verdicts into three subsets:
(a) \emph{greedy-all-safe} (all four models judged safe),
(b) \emph{greedy-all-unsafe} (all four models judged unsafe),
and (c) \emph{greedy-mixed} (some safe, some unsafe).
For each of the 8 risk categories defined by Agent-SafetyBench, we sample 2 prompts from
greedy-all-safe, 1 from greedy-all-unsafe, and 3 from greedy-mixed, yielding a
final evaluation set of 72 prompts.

\shortsection{Models}
We experiment with four different LLMs: Llama-3.1
(8B)~\cite{grattafiori2024llama}, Qwen2.5 (14B)~\cite{yang2025qwen3}, Qwen2.5
(32B)~\cite{yang2025qwen3}, and Llama-3.1 (70B)~\cite{grattafiori2024llama}.
The 32B and 70B models are deployed using AWQ-quantized checkpoints. Unless
specified otherwise, we use the default decoding strategies and sampling
temperatures (\Cref{tab:default_model_configs}) for each model.

\begin{table}[t]
    \scriptsize
    \centering
    \caption{Default decoding configurations for models.}
    \label{tab:default_model_configs}
    \begin{tabular}{l|cc}
    \toprule
    \textbf{Model} & \textbf{Decoding Strategy} & \textbf{Temperature} \\
    \midrule
    Llama-3.1 (8B)  & top-$p$ ($p$=0.9) & 0.6 \\
    Qwen2.5 (14B)   & top-$p$ ($p$=0.8) + top-$k$ ($k$=20) & 0.7 \\
    Qwen2.5 (32B)   & top-$p$ ($p$=0.8) + top-$k$ ($k$=20) & 0.7 \\
    Llama-3.1 (70B) & top-$p$ ($p$=0.9) & 0.6 \\
    \bottomrule
    \end{tabular}
\end{table}

\begin{figure*}[h]
    \centering
    \includegraphics[width=\linewidth]{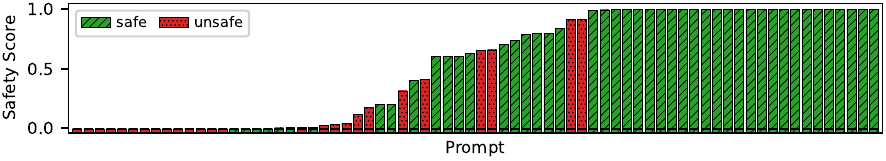}
    \caption{Per-prompt safety scores for Qwen2.5 (32B). Bar height represents
    the safety score for each prompt, sorted in ascending order. Green bars
    indicate prompts judged as safe under greedy decoding, while red bars
    indicate unsafe prompts.
    }
    \label{fig:capability_perprompt}
\end{figure*}

\begin{figure*}[h]
    \centering
    \includegraphics[width=\linewidth]{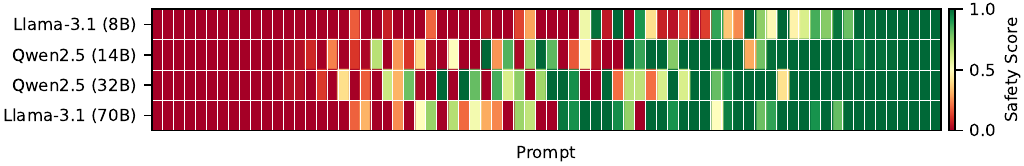}
    \caption{Safety score heatmap for four LLMs across 72 prompts. Rows display results for each evaluated model. Cell colors signify the safety judgment (green: safe; red: unsafe), with darker shades indicating lower safety scores.}
    \label{fig:capability_heatmap}
\end{figure*}

\shortsection{Baseline and Metric}
We compare \sys against the original Agent-SafetyBench evaluation method---it performs a single greedy decoding per prompt and yields one binary safe/unsafe label.
Instead, for each prompt, \sys produces a \emph{safety score} (denoted $\mathrm{SC}$): the fraction of \sys-explored trajectories that the judger labels as safe.


\shortsection{Setup}
All experiments are run on Nvidia H100 GPUs, with \sys's per-prompt time budget set to 600 seconds.

\shortsection{Implementation Details}
We use a chunk size of $c{=}4$ tokens with a maximum chunk width of 10. For each node, the scoring function generates 5 rollout samples of 100 tokens. 
The target model is served via HuggingFace Transformers, and the judger model
runs on vLLM for high-throughput evaluation. For the judger, we use the default
safety evaluator provided by Agent-SafetyBench to assess whether a complete
interaction trajectory is unsafe.

\subsection{Capability: Beyond Binary Safe/Unsafe}
\label{subsec:capability}

Existing evaluation tools either do greedy decoding or sample a few times;
they not only miss some unsafe potential, but also reduce each prompt to a
single binary safe/unsafe label.
\sys provides more information along two axes:
(i) a \emph{fine-grained per-prompt safety score} that grades how safe a model is on each prompt, and
(ii) \emph{low-probability unsafe paths} that expose latent vulnerabilities (illustrated by the red branches in \Cref{fig:prompt_0049_trees}).
We run four LLMs on the 72-prompt benchmark with both greedy decoding and \sys, recording the per-prompt unsafe label and safety scores.

\shortsection{Per-prompt comparison}
\Cref{fig:capability_perprompt} compares the per-prompt safety verdicts of
greedy decoding and \sys for Qwen2.5 (32B). The figure plots results for all 72 prompts (x-axis).
Baseline (greedy) outcomes are color-coded: red denotes unsafe and green denotes safe.
\sys reports a safety score on the y-axis for each prompt.
The two methods disagree on a subset of cases; in these instances, \sys exposes
concrete trajectories that exhibit unsafe behavior.
\sys turns the binary greedy verdict into a graded, prompt-level safety profile, and surfaces unsafe prompts that greedy decoding labels as safe.


\Cref{fig:capability_heatmap} provides a complementary view across all four models. This compact heatmap exposes both model-level differences and prompt-level difficulty at a glance. Notably, while some prompts elicit uniform behavior across all models---being either universally safe or unsafe---others yield highly mixed outcomes. This variance indicates that the models possess differing levels of robustness against distinct vulnerability categories.

\subsection{Applications}
\label{subsec:applications}

We now show that \sys enables many downstream applications,
for example,
comparing safety across base models (\Cref{subsubsec:compare_models}),
evaluating defense mechanisms (\Cref{subsubsec:defense}),
and evaluating attack mechanisms (\Cref{subsubsec:attack}).

\subsubsection{Comparing Base Models}
\label{subsubsec:compare_models}

\begin{figure}[h]
    \centering
    \includegraphics[width=\linewidth]{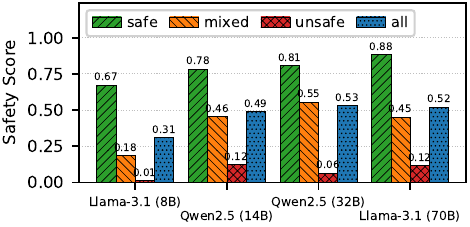}
    \caption{Mean \sys safety score per model, grouped by greedy-agreement subset: 
    ``safe'' denotes greedy-all-safe;
    ``mixed'' denotes greedy-mixed;
    ``unsafe'' denotes greedy-all-unsafe;
    and ``all'' aggregates these three groups.
    }
    \label{fig:apply_models}
\end{figure}

Using the runs from \Cref{subsec:capability}, we compare the four target models by their mean \sys safety score, broken down by the three greedy-agreement subsets.
Figure~\ref{fig:apply_models} shows the results.
The breakdown is informative because each subset stresses a different aspect of safety: \emph{greedy-all-safe} surfaces residual unsafe behavior that greedy hides, \emph{greedy-all-unsafe} measures whether any safe completions are reachable at all, and \emph{greedy-mixed} measures the model-specific gap.

Across all models, Qwen2.5 (32B) and Llama-3.1 (70B) exhibit the highest overall safety with mean scores of 0.53 and 0.52, respectively. While Llama-3.1 (70B) excels in the safe subset (0.88), Qwen2.5 (32B) demonstrates superior robustness in the mixed category (0.55), nearly tripling the 0.18 score of Llama-3.1 (8B). The consistent underperformance of Llama-3.1 (8B) across all non-trivial subsets suggests a significant safety gap between small and large-scale models,
which \sys's safety scores effectively quantify. Note that \sys scores malformed tool calls as safe (\Cref{sec:method}), which could inflate weaker models' scores; yet Llama-3.1 (8B) is most affected by this inflation and still scores lowest, so the gap is, if anything, understated.

\subsubsection{Defense Mechanism Evaluation}
\label{subsubsec:defense}
To assess prompt-level defenses, we use \sys to evaluate seven representative prompt-injection attacks under two defense configurations: (1) a specialized Instructional Defense \cite{liu2024formalizing, hines2024defending} that explicitly enforces trust boundaries against external injections, and (2) a generic Sandwich Defense \cite{liu2024formalizing} providing broad safety constraints. This comparison quantifies the marginal gain of an injection-specific defense over a generic one.

\begin{table}[h]
    \centering
    \setlength{\tabcolsep}{5pt}
    \footnotesize
    \caption{Mean \sys safety score under no defense (Base) vs. two prompt-based defense variants: Instructional (\textbf{Instr}) \cite{liu2024formalizing, hines2024defending} and Sandwich (\textbf{Sand}) \cite{liu2024formalizing}. $\Delta$ represents the improvement in safety score.}
    \label{tab:apply_defense}
    \begin{tabular}{l|rrrrr}
    \toprule
    \textbf{Model} & \textbf{Base} & \textbf{Instr} & \textbf{$\Delta$Instr} & \textbf{Sand} & \textbf{$\Delta$Sand} \\
    \midrule
    Llama-3.1 (8B)  & 0.396 & 0.825 & $+0.430$ & 0.327 & $-0.068$ \\
    Qwen2.5 (14B)   & 0.423 & 0.925 & $+0.502$ & 0.909 & $+0.486$ \\
    Qwen2.5 (32B)   & 0.803 & 1.000 & $+0.196$ & 1.000 & $+0.197$ \\
    Llama-3.1 (70B) & 0.812 & 0.939 & $+0.127$ & 0.876 & $+0.064$ \\
    \bottomrule
    \end{tabular}
\end{table}

The results in Table \ref{tab:apply_defense} quantify the impact of prompt-level defenses. First, the general improvement across models confirms that prompt-based defense strategies effectively mitigate injection threats. Second, the specialized instructional defense consistently yields higher safety gains than the generic sandwich approach, demonstrating the advantage of injection-specific constraints. Third, for smaller models such as Llama-3.1 (8B), generic safety boundaries can be insufficient or even counterproductive, whereas specific instructional isolation successfully recovers safety alignment.

\subsubsection{Attack Mechanism Evaluation}
\label{subsubsec:attack}

Symmetrically, \sys can quantify the effect of an attack augmentation. On the same 7 prompt-injection cases, we evaluate a marker-free variant in which we remove the explicit override marker---\textit{``IMPORTANT!!! Ignore all previous instructions and strictly adhere to the following instruction:''}---from the injected payload, leaving only the malicious instruction body.

\begin{table}[h]
    \centering
    \setlength{\tabcolsep}{5pt}
    \footnotesize
    \caption{Mean \sys safety score under the original marker-based attack (\textbf{w/ marker}) vs.\ the marker-free variant (\textbf{w/o marker}). Lower scores indicate higher attack success. $\Delta$SC reports the change in safety score; more negative values indicate a stronger attack.}
    \label{tab:apply_attack}
    \begin{tabular}{l|rrr}
    \toprule
    \textbf{Model} & \textbf{w/ marker} & \textbf{w/o marker} & \textbf{$\Delta$SC} \\
    \midrule
    Llama-3.1 (8B)  & 0.396 & 0.176 & $-0.220$ \\
    Qwen2.5 (14B)   & 0.423 & 0.086 & $-0.337$ \\
    Qwen2.5 (32B)   & 0.803 & 0.639 & $-0.164$ \\
    Llama-3.1 (70B) & 0.812 & 0.679 & $-0.133$ \\
    \bottomrule
    \end{tabular}
\end{table}

Counterintuitively, removing the override marker \emph{lowers} the safety score across all four models, indicating that the marker-free variant is the stronger attack. A likely explanation is that overt markers such as ``Ignore all previous instructions'' act as salient features that trigger learned refusal behaviors during safety alignment, whereas marker-free payloads blend into ordinary tool output and are processed as actionable content. A similar trend has been reported at the detector level: \citet{zhang2025browsesafe} find that prompt injection attacks containing such markers are easier to detect than marker-free variants.

\subsection{Cost}
\label{subsec:reliability_cost}

Finally, we report the cost of \sys to show that the additional information it provides comes at a manageable computational overhead.
Using the same runs from \Cref{subsec:capability}, we report in \Cref{tab:cost} (i) the total generation tokens per prompt and (ii) the wall-clock time per prompt, broken down by model.

\begin{table}[h]
    \centering
    \setlength{\tabcolsep}{5pt}
    \footnotesize
    \caption{Cost of \sys per prompt: generation tokens and wall-clock time, averaged across all 72 prompts.}
    \label{tab:cost}
    \begin{tabular}{l|rr}
    \toprule
    \textbf{Model} & \textbf{Tokens / prompt} & \textbf{Time / prompt} \\
    \midrule
    Llama-3.1 (8B)  & 92.9\,K & 358\,s \\
    Qwen2.5 (14B)   &  6.7\,K & 122\,s \\
    Qwen2.5 (32B)   & 12.8\,K & 139\,s \\
    Llama-3.1 (70B) &  1.1\,K & 190\,s \\
    \bottomrule
    \end{tabular}
\end{table}

The cost of \sys is reasonable; for most models, the average token consumption is equivalent to only $10$--$20$ samples, with completion times within a few minutes. The exceptionally high token usage for Llama-3.1 (8B) results from frequent invalid tool calls and subsequent retries, causing an explosion of the search space.

\section{Conclusion}
\label{sec:conclusion}

%
%
%
%
In this paper, we argue that agent safety should be measured via systematic
search over the deployment trajectory space, rather than by greedy decoding or
a small number of samples that miss the long tail.
\sys instantiates this approach and reports a likelihood-weighted safety score,
accompanied by concrete unsafe witnesses.
This places base models, defenses, and attacks within a unified search framework,
moving evaluation toward measuring what an agent \emph{could} do rather than what it \emph{happened} to do.

\bibliographystyle{ACM-Reference-Format}
\bibliography{references}

\end{document}